\documentclass[atoms,article,moreauthors,pdftex,10pt,a4paper]{mdpi} 
\firstpage{1} %%MDPI internal note: new layout%%
\articlenumber{0000}
\doinum{10.3390/------}
\pubvolume{xx}
\pubyear{2016}
\copyrightyear{2016}
\externaleditor{Academic Editor: name}
\history{Received: date ; Accepted: date ; Published: date}
\newcommand{\be}{\begin{equation}}
\newcommand{\ee}{\end{equation}}
\newcommand{\ket}[1]{\left| #1 \right\rangle}
\newcommand{\sub}[1]{_{\mbox{\scriptsize #1}}}
\renewcommand{\vec}[1]{{\bf #1}}
\newcommand{\ehat}{\hat{\epsilon}}
\newcommand{\bhat}{\hat{b}}

 \theoremstyle{mdpi}
 \newcounter{thm}
 \newcounter{ex}
 \newcounter{re}
 \theoremstyle{mdpidefinition}

\Title{Obtaining Atomic Matrix Elements from Vector Tune-Out Wavelengths using Atom Interferometry}

\Author{A. J. Fallon$^{1}$ and C. A. Sackett $^{1}$*}
\address[1]{%
$^{1}$ \quad Physics Department, University of Virginia, Charlottesville, VA 22904, USA}

\corres{Correspondence: sackett@virginia.edu}

\abstract{
Accurate values for atomic dipole matrix elements are useful in many areas 
of physics, and in particular for interpreting experiments such as atomic 
parity violation.
Obtaining accurate matrix element values is a challenge for both 
experiment and theory. A new technique that can be applied to this 
problem is tune-out spectroscopy, which is the measurement of 
light wavelengths where the electric polarizability of an atom has a zero. 
Using atom interferometry methods, tune-out wavelengths can be 
measured very accurately. Their values depend on the ratios of 
various dipole matrix elements and are thus useful for constraining theory and 
broadening the application of experimental values. Tune-out wavelength 
measurements to date have focused on zeros of the scalar polarizability, 
but in general the vector polarizability also contributes. We show here 
that combined measurements of the vector and scalar polarizabilities can 
provide more detailed information about the matrix element ratios, and 
in particular can distinguish small contributions from the atomic core 
and the valence tail states. These small contributions are 
the leading error sources in current parity violation calculations for cesium.
}

\keyword{atom interferometry; atomic matrix elements; tune-out wavelength; polarizability}

\begin{document}

\section{Introduction}

Most of our knowledge about atoms derives from spectroscopic studies. 
Conventional spectroscopy provides precise information about the energy of
electronic states in an atom. These states can be represented as poles in 
the frequency response of the atom to an applied field. 
A complementary approach can be found in the 
recently developed technique of tune-out spectroscopy, in which
zeros of the atomic frequency response are 
measured \cite{LeBlanc07,Arora11}.
Although tune-out measurements are generally more complicated to implement
than conventional spectroscopy, it is still possible to achieve high
accuracy results \cite{Holmgren12,Herold12,Henson15,Leonard15,Schmidt16}.

An important feature of tune-out spectroscopy is that it provides information
about the relationship between the atomic response of different states.
For example, in between a nearby pair of dipole-allowed transitions there is
a zero in the dynamic electric polarizability, where the positive polarizability
from one state perfectly cancels the negative polarizability from the other
state.  The location of the zero
depends primarily on the ratio of the electric dipole matrix elements
of the two states.  While there are techniques to directly measure
matrix elements for some states, 
the ratio determined via tune-out spectroscopy can be
much more accurate than the ratio of direct measurements.  For instance,
our recent tune-out frequency
measurement for the rubidium $5P$ states improved the accuracy
of the matrix element ratio by a factor of 100 compared to 
direct measurements  \cite{Leonard15}.
Furthermore, for many states
it is difficult to measure the matrix elements directly
with good precision. Tune-out spectroscopy can be used in such cases to relate
the desired matrix element to that of a more easily measured transition.
For instance, Herold {\em et al.}\ improved knowledge of the $6P$ matrix
elements in rubidium by a factor of ten by relating them to the better-known
$5P$ elements \cite{Herold12}.

While these results demonstrate the utility of tune-out spectroscopy,
a challenge is that to some degree the frequency of any single response zero
depends on all of the accessible states and matrix elements in the atom
\cite{Arora11}. 
To deal with this,
theoretical estimates are used for contributions that are not
of direct interest. This introduces additional sources of uncertainty
and limits the applicability of the measurements to systems where 
high-quality theoretical estimates are available.

We present here a technique to reduce this dependence on theory.
Up to now, most studies have centered on the scalar response of the atom,
which can be obtained by averaging over the atomic spin states and/or the
optical polarization of the light. 
However, spin-polarized atoms also have 
a strong vector response, which can be measured by varying
the light polarization and the spin orientation. 
The effect on the tune-out wavelength
was measured in a recent study
by Schmidt {\em et al.} \cite{Schmidt16}

Like the scalar response, the
vector response depends on multiple transition matrix elements.
However, these elements combine in a different way for the scalar and
vector quantities.  We show here that by
making joint measurements of both responses, different contributions
to the electric polarizability can be experimentally resolved. 
This can lead to improved accuracy and can provide experimental information
about matrix elements that cannot easily be observed directly.

Of particular significance, it is
possible to independently determine
the effects of the atomic core and of the infinite
manifold of high-lying valence states.
The electric polarizability of the ionic core can in some cases
be determined directly by Rydberg atom experiments \cite{Mayer33}, but such
a measurement
requires a theoretical correction to account for core-valence interactions
in the ground-state atom \cite{Safronova11}. 
The `tail' contribution from the valence manifold is even more challenging,
and no direct measurement of its effect seems possible due to the many states
involved \cite{Dzuba12b}.

These core and tail contributions are small but 
have significant importance. In particular, the well
known experimental measurements of parity violation in cesium \cite{Wood97}
can be related to fundamental quantities in high-energy physics,
but this requires precise knowledge of atomic dipole matrix elements
including the core and tail contributions. Theoretical uncertainty
in the tail is currently the dominant source of error in interpreting
the experimental results \cite{Dzuba12b}. Experimental measurement of the tail
contribution using the tune-out spectroscopy
technique discussed here could therefore be of great utility.

Accurate knowledge of atomic matrix elements is useful
for other applications as well. Important examples include the
prediction and characterization of Feshbach resonances in atomic collisions
\cite{vanKempen02,Claussen03} and estimation of blackbody shifts for atomic
clocks \cite{Sherman12,Safronova12}. 

This paper presents calculations regarding
the utility of vector tune-out measurements in
alkali atoms, with rubidium as a specific example. Section 2
presents an analysis of the lowest tune-out frequency,
while Section 3 shows that more information can be obtained
by combining measurements at several tune-out frequencies. Section 4
discusses experimental considerations, and Section 5 offers conclusions.

\section{Vector tune-out analysis}

The response of an atom to an off-resonant optical field at frequency $\omega$
is largely governed by the
electric polarizability, $\alpha(\omega)$. For an alkali atom in state
$i$, this can be expressed as \cite{Mitroy10}
\be
\label{polz1}
\alpha(\omega) = \frac{1}{\hbar} \sum_f 
\frac{2\omega_{if}}{\omega_{if}^2-\omega^2} \left| d_{if}\right|^2
+\alpha_c + \alpha_{cv}.
\ee
The sum is over all excited states $f$ of the valence electron,
and $\omega_{if}$ is the transition frequency between $i$ and $f$.
Here we neglect hyperfine structure and let
$i$ and $f$ represent individual Zeeman states
$\ket{i} = \ket{n\, L\, J\, m_J}$ and 
$\ket{f} = \ket{n'\, L'\, J'\,  m_J'}$.
The matrix element
$d_{if}$ is $\langle f | \vec{d}\cdot\ehat | i \rangle$
where $\vec{d}$ is the dipole operator and $\ehat$ is the polarization
vector of the light. Given the $S$ ground state for alkali atoms, 
only excited $P$ states
will appear in the sum. The polarizability contribution of the core electrons
is represented by $\alpha_c$, while $\alpha_{cv}$ is the correction
accounting for core-valence interactions \cite{Arora11}. 

Using Eq.~\eqref{polz1}, 
the interaction energy of the atom with the field can be 
expressed as
\be
\label{energy}
U = -\frac{1}{2} \alpha \langle \mathcal{E}^2 \rangle
\ee
where $\mathcal{E}$ is the electric field of the light and the 
angle brackets denote a time average. 

It is often convenient to decompose Eq.~\eqref{polz1} into
spherical tensor components. The interaction energy can then be
expressed as \cite{Kien13}
\be
U = -\frac{1}{2} \langle \mathcal{E}^2 \rangle \left[
\alpha^{(0)} - \frac{i}{2} 
(\ehat^*\times \ehat)\cdot\bhat \, \frac{m_J}{J} \alpha^{(1)}
\right]
\ee
where the $\alpha^{(0)}$ is the scalar component and $\alpha^{(1)}$ is 
the vector component.
The quantization axis for the states
is defined by a magnetic field pointing in the $\bhat$ direction.
The polarizability components themselves can be calculated 
in terms of reduced matrix elements $D_{if}$ as
\be
\label{alpha0}
\alpha^{(0)} = \frac{1}{3\hbar} \sum_f |D_{if}|^2
\frac{ \omega_{if}}{\omega_{if}^2 - \omega^2} + \alpha_c + \alpha_{cv}^{(0)}
\ee
and
\be
\label{alpha1}
\alpha^{(1)} = \frac{1}{3\hbar} \sum_f 
C_{J'} |D_{if}|^2
\frac{ \omega}{\omega_{if}^2 - \omega^2} +\alpha_{cv}^{(1)}.
\ee
Here the sums over excited states $f$ include $n'$ and $J'$ but not $m_J'$.
The reduced matrix elements
$D_{if} \equiv \langle 5S_{1/2} || d ||\, n'P_{J'} \rangle$
are defined
using the convention of, for instance, Ref.~\cite{Arora11}. In
$\alpha^{(1)}$, $C_{J'} = 3J'-7/2$ 
is either -2 or +1 depending on the excited state angular momentum.
Since the ionic core has zero spin, the core term
$\alpha_c$ has only a scalar contribution. The core-valence term
$\alpha_{cv}$ can have a vector component as well, denoted by
$\alpha_{cv}^{(1)}$. 

High precision measurements require that hyperfine structure be taken
into account, resulting in more complicated formulas for the $\alpha^{(q)}$
components \cite{Kien13}. 
However, this will not produce any
qualitative changes to the results described here.

Tune-out spectroscopy locates frequencies $\omega_0$ where $\alpha = 0$. 
Our concern here is to relate the value of $\omega_0$ to the 
dipole matrix elements and core contributions. We
consider first the lowest tune-out frequency for Rb atoms, located in
between the $5S_{1/2}\rightarrow 5P_{1/2}$ transition at 795 nm and 
the $5S_{1/2}\rightarrow 5P_{3/2}$ transition at 780 nm. 
The atoms are initially spin polarized with $m_J = +1/2$.

Figure 1 shows the scalar and vector polarizabilities in this
wavelength range.
The atomic parameters used for the calculation are summarized
in Tables 1 and 2, and will be discussed further below.
The tune-out frequency $\omega_0$ is the solution of 
$\alpha^{(0)}-v\alpha^{(1)} = 0$,
where $v \equiv (i/2)(\ehat^*\times \ehat)\cdot\bhat$ 
can vary from -1/2 to +1/2 as the light
polarization and magnetic field direction are adjusted. The data points
in Fig. 2(a) plot the wavelengths $\lambda_0$ corresponding to 
numerical solutions for $\omega_0$ at different $v$. 
The dominant behavior is clearly linear, which can be understood
as follows:

\begin{figure}[H]

\centering
\includegraphics[width=3in]{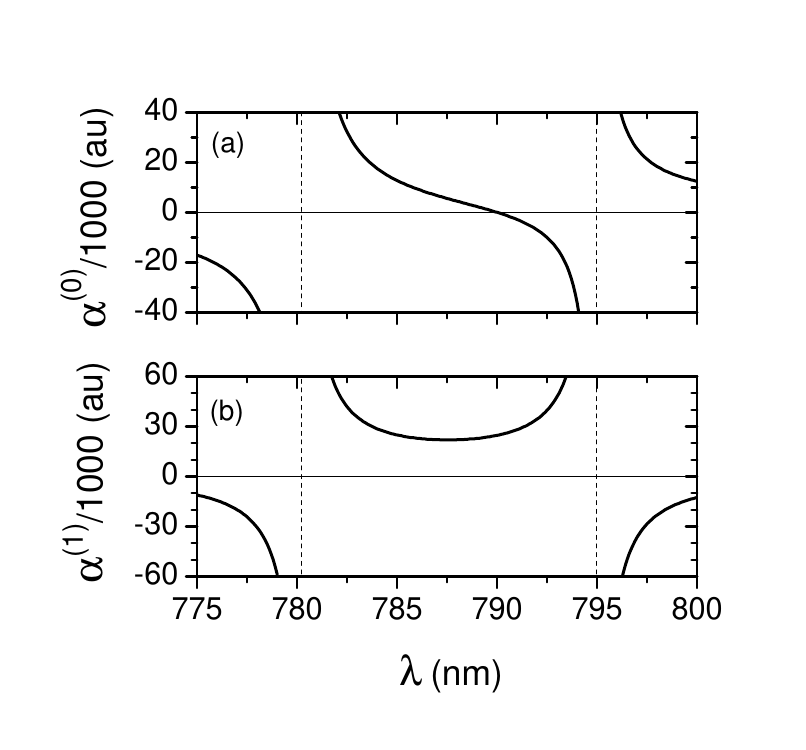}
\caption{Scalar polarizability $\alpha^{(0)}$ and vector polarizability
$\alpha^{(1)}$ for Rb atoms near a wavelength $\lambda$ of 790 nm.
Values plotted are divided by 1000 and given in atomic units. The
dashed lines indicate the locations of the $5P$ resonances,
with $5P_{1/2}$ at 795 nm and $5P_{3/2}$ at 780 nm. The scalar
polarizability crosses zero near 790 nm, defining a tune-out 
wavelength $\lambda_0$.
}
\label{fig1}
\end{figure}

In this frequency range, the two $5P$ transitions dominate the sums in
Eqs.~(\ref{alpha0}) and (\ref{alpha1}). In the limit that the fine structure
splitting is small compared
to $\omega$, 
it is reasonable to keep only the $5P$ terms and approximate
them as
\be
\label{linear}
\alpha \approx \frac{1}{6} 
\left( \frac{\left|D_1\right|^2 (1-2v)}{\omega_{1/2}-\omega}
+ \frac{\left|D_2\right|^2(1+v)}{\omega_{3/2}-\omega}\right),
\ee
where we abbreviate here $D_1$ for $D_{5S_{1/2},5P_{1/2}}$, 
$D_2$ for $D_{5S_{1/2},5P_{3/2}}$, 
and $\omega_J$ for $\omega_{5S_{1/2},5P_J}$.
We can also approximate $|D_2|^2 \approx 2 |D_1|^2$, 
as the state degeneracies would suggest.
The root of Eq.~\eqref{linear} is then easily found as 
\be
\label{omega0a}
\omega_{a} = \omega_{1/2} + \frac{\Delta}{3} + \frac{2\Delta}{3}v,
\ee
for $\Delta = \omega_{3/2}-\omega_{1/2}$.
The corresponding wavelength is plotted as the line 
in Fig. 2(a), and
evidently accounts for the main features of the exact solution.
At this level, no useful information is obtained from the tune-out measurement,
since the fine structure splitting is already known from 
conventional spectroscopy.

\begin{figure}
\includegraphics[width=4in]{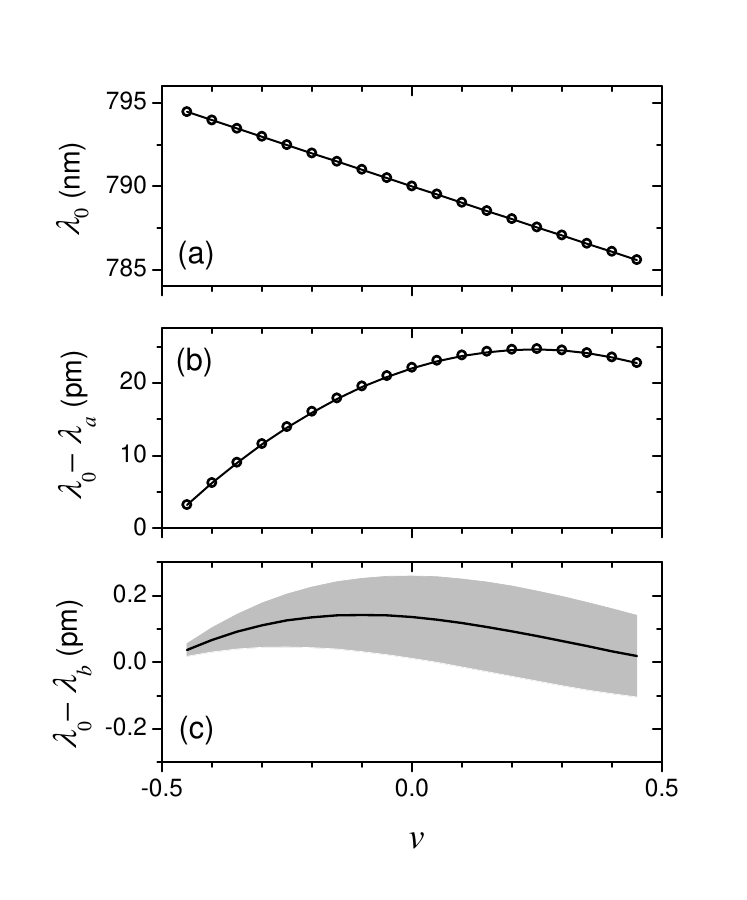}
\caption{Tune-out wavelength near 790 nm for Rb atoms. (a) 
Variation of tune-out wavelength with polarization parameter 
$v= i(\ehat^*\times\ehat)\cdot\bhat/2$ for light polarization
vector $\ehat$ and quantization axis direction $\bhat$. Atoms
are taken to be in the spin-polarized ground state $m_J = 1/2$. Data
points show the numerical solution $\lambda_0$ 
for the zero of the polarizability
$\alpha(\omega) = \alpha^{(0)}(\omega) - v\alpha^{(1)}(\omega)$
from the full model of Eqs.~\protect\eqref{alpha0} 
and \protect\eqref{alpha1}. The
straight line shows the approximate solution $\lambda_a$ from
Eq.~\protect\eqref{omega0a}. (b) Deviation from linearity $\lambda_0 -
\lambda_a$.
Points show the difference between the full numerical solution
and the approximation of Eq.~\protect\eqref{omega0a}. The solid curve shows 
$\lambda_b-\lambda_a$ for the 
polynomial approximation $\lambda_b$ from 
Eq.~\protect\eqref{polyv}.
(c) The black curve shows the deviation $\lambda_0-\lambda_b$. 
The gray band illustrates the variation obtained in $\lambda_0$ as the 
model parameters $R_{5,3/2}$, $\alpha_c+\alpha_{cv}^{(0)}$, 
$|t_{1/2}|^2$ and $|t_{3/2}|^2$ (see Table 2)
are varied by their uncertainties.
}
\end{figure}

However, the deviations between the linear approximation 
and the exact solution are significant,
as seen in Fig.~2(b). These deviations can be 
compared to the 0.03-pm accuracy of
the experimental measurements in \cite{Leonard15}. To analyze these deviations
we develop a more accurate approximation.
We retain the $5P$ states as the dominant terms, 
and in particular assume that the frequency variation
of the other terms in Eqs.~\eqref{alpha0} and \eqref{alpha1} 
can be neglected since those states are
far from resonance. We also allow for a non-ideal matrix 
element ratio as $|D_2/D_1|^2 = 2 + \delta R$. 
The polarizability can then be expressed as
\be
\alpha \approx \frac{1}{3} \left|D_1^2\right| \left[ F(\omega) + A - Bv + 
\delta R \frac{\omega_{3/2}-v\omega}{\omega_{3/2}^2-\omega^2}\right]
\ee
for
\be
F(\omega) = \frac{\omega_{1/2}+2v\omega}{\omega_{1/2}^2-\omega^2} 
+ \frac{2(\omega_{3/2}-v\omega)}{\omega_{3/2}^2-\omega^2},
\ee
\be
A = \frac{1}{\left|D_1\right|^2}\left[3\alpha_c + 3\alpha_{cv}^{(0)} + 
\sum_f \left|D_{if}\right|^2 \frac{\omega_{if}}{\omega_{if}^2-\omega^2}\right],
\ee
and
\be
\label{B}
B = \frac{1}{\left|D_1\right|^2}\left[3\alpha_{cv}^{(1)} + 
\sum_f \left|D_{if}\right|^2 C_{J'} \frac{\omega}{\omega_{if}^2-\omega^2}\right].
\ee
We use $D_1 = 4.2339(16)$ au as a common normalizing factor
\cite{Volz96,Leonard15}.
The sums in $A$ and $B$ are over valence states 
with $n'>5$. We evaluate these terms 
at $\omega_{a0} \equiv \omega_{1/2} + \Delta/3$.
Since the $\delta R$ term includes a small denominator, we approximate it 
more accurately using $\omega_a(v)$ from \eqref{omega0a}. This gives
\be
\frac{\omega_{3/2}-v\omega}{\omega_{3/2}^2-\omega^2} \approx \frac{3}{4\Delta},
\ee
independent of $v$ to lowest order.

The solution of $F(\omega) = 0$
can be obtained analytically as $\omega_{F}$, and then the 
effect of the small terms give an
approximate root
\be
\omega_b = \omega_{F} - 
%\left.\left(\frac{dF}{d\omega}\right)^{-1}\right|_{\omega=\omega_{0F}}
\left(\frac{dF}{d\omega}\right)^{-1}
\left(A-vB+\frac{3}{4\Delta} \delta R\right)
\ee
with the derivative evaluated at $\omega_{F}$.
Expanding this to second order in the fine-structure splitting 
$\Delta$ and to first order in the small parameters
$A$, $B$ and $\delta R$ yields a cubic polynomial
in $v$,
\be
\label{polyv}
\omega_b = \sum_n b_n v^n
\ee
with
\be
b_0 = \omega_{1/2} + 
\frac{1}{3}\Delta - \frac{1}{9}\frac{\Delta^2}{\omega_{1/2}}
- \frac{4}{27} \Delta^2 \left(A + \frac{3\delta R}{4\Delta} \right),
\ee
\be
\label{a1}
b_1 = \frac{2}{3}\Delta - \frac{1}{9}\frac{\Delta^2}{\omega_{1/2}}-
\frac{4}{27} \Delta^2 \left(A + \frac{3\delta R}{4\Delta}\right)
+\frac{4}{27} \Delta^2 B ,
\ee
\be
b_2 = \frac{2}{9} \frac{\Delta^2}{\omega_{1/2}} + \frac{8}{27}\Delta^2 
\left( A + \frac{3\delta R}{4\Delta}\right)
+\frac{4}{27}\Delta^2 B 
\ee
and
\be
b_3 = - \frac{8}{27} \Delta^2 B.
\ee
This approximate solution is shown as the curve in Fig~ 2(b),
and agrees well with the full numerical results. The difference between
the polynomial approximation and the full solution is shown in Fig.~3(c),
where the shaded area represents the uncertainty in the theoretical
parameters of the model.

It is observed that the polynomial approximation 
$\omega_b(v)$ depends on the non-$5P$ state parameters only through 
the combinations $A + 3\delta R/4\Delta$
and $B$. The first combination can be obtained anyway from a scalar
measurement with $v=0$, so $B$ is of more interest here.
The sum in Eq.~\eqref{B} can be broken into a finite number of
terms from $n'=6$ to some $n\sub{max}$, plus an infinite sum over
states very near the dissociation limit. It is possible to obtain
accurate theoretical estimates for the matrix elements in the finite
terms. For instance, Table 1 shows results up to $n\sub{max} = 12$
taken from Ref.~\cite{Leonard15}, which together provide a contribution to 
$B$ of $0.029 \pm 0.002$ au. 

For the remaining tail contribution,
theoretical estimates are uncertain
due to significant dependence on the calculational methods used. The 
estimates used in Ref.~\cite{Leonard15} were
\be
\label{TJ}
T_{J'} \equiv \frac{1}{3} 
\frac{\omega\sub{ion}}{\omega\sub{ion}^2-\omega_{a0}^2} 
\sum_{n'>12} |D_{n',J'}|^2
\ee
with $T_{1/2} \approx 0.022$ au and $T_{3/2} 
\approx 0.075$ au, both with error estimates comparable
to their values.
In \eqref{TJ}, $\omega\sub{ion}$ corresponds to the ionization 
frequency at a wavelength of 
297.8 nm. The contribution of these terms to $B$ will be
$(\omega_{a0}/\omega\sub{ion})(T_{3/2}-2T_{1/2}) \approx
0.002 \pm 0.007$ au.

We are not aware of any previous calculations of the 
vector core contribution $\alpha_{cv}^{(1)}$. The scalar
term $\alpha_{cv}^{(0)}$ is calculated to be $-0.37 \pm 0.04$ au
\cite{Safronova_private}.
From the general form of the polarizability expansion,
we expect the vector term to be smaller by a factor on the order
of $\omega_{a0}/\omega\sub{core}$ where $\omega\sub{core}$ is
the lowest excitation frequency of the ionic core. For
Rb$^+$ this lies at a wavelength of 74 nm, so we expect
$\alpha_{cv}^{(1)} \sim -0.04$ au. If we estimate the uncertainty
as comparable to this value, we obtain a total estimate
for $B$ of $0.025\pm 0.016$ au, with uncertainty dominated by
the tail and core terms. A precise measurement of $B$ would
therefore provide an experimental constraint on these
uncertain quantities.

To estimate the feasible measurement accuracy, we use
the error estimates discussed in Ref.~\cite{Leonard15}.
The reported wavelength error $\delta\lambda_0 \approx 0.03$ pm
corresponds to $\delta\omega_0 \approx 2\times 10^{-9}$
au. Considering, for instance, the linear term
$a_1$ from \eqref{a1} and assuming that $v$ can
be varied by about one, we get an expected
accuracy
\be
\delta B \approx \frac{27}{4} \frac{\delta\omega_0}{\Delta^2}
\approx 0.01~\text{au}.
\ee
Since this is smaller than the theoretical uncertainty,
it can be expected that experimental
measurements will provide useful information.

\section{Multiple tune-out frequency analysis}

The previous discussion illustrates that measurement of
vector tune-out wavelengths
can provide an experimental constraint on parameters of theoretical
interest, thanks to the differing character of their polarization
dependence. However, the measurements near 790 nm cannot
provide definitive values for these parameters. 
We show now that by combining vector measurements
around multiple tune-out frequencies, we can obtain
more direct results. This is possible because the different components
have different frequency dependence, in addition to the polarization
dependence. By varying both the frequency and polarization, enough
information can be obtained to extract the values of individual contributions.

After the tune-out wavelength at 790 nm, the next longest tune-out wavelengths
in Rb are a pair associated with the $6P$ states near 420 nm. 
The scalar zeros were measured by Herold {\em et al.}\ 
to be 423.018(7) nm and 421.075(2) nm. We show that vector
measurements near these wavelengths are sufficient to extract the 
polarizability terms of interest. Another pair of tune-out wavelengths
near 360 nm is associated with the 7P states, and is also reasonably
accessible.

The polarizabilities near 420 nm can be calculated using the
same equations \eqref{alpha0} and \eqref{alpha1}. To good approximation,
the core terms should be the same as at 790 nm, since $\omega\sub{core}$ is
still much larger than $\omega$ \cite{Mitroy10}. Figure 3 shows plots of
the tune-out wavelengths as functions of $v$. The behavior
is more complicated here since there are comparable contributions from 
several states, and an algebraic analysis involves solutions to
high-order polynomials.
So, rather than developing an analytical model, we
consider a numerical fit to artificially generated synthetic data.

\begin{figure}
\includegraphics[width=4in]{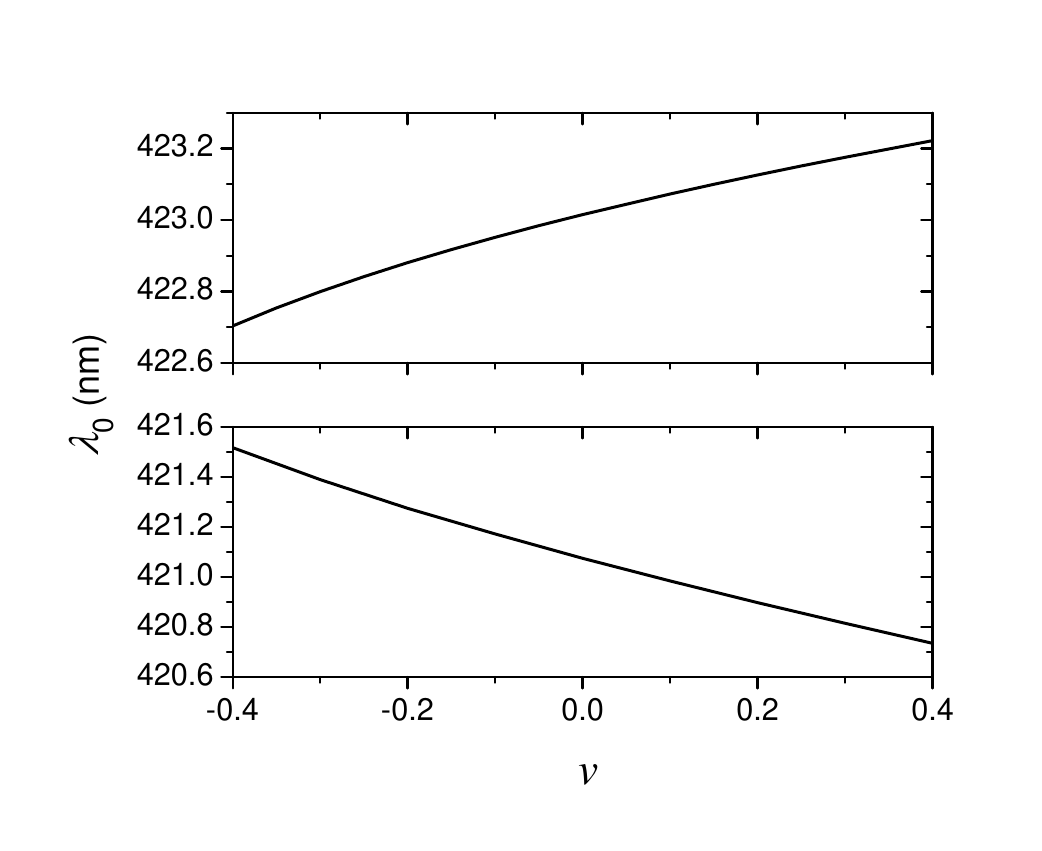}
\caption{
Tune-out wavelengths $\lambda_0$ for Rb atoms near 420 nm. For
each value of the polarization parameter $v$, there are two 
tune-out wavelengths near the $6P_{1/2}$ and $6P_{3/2}$ states.
}
\end{figure}

To generate a synthetic data set, we calculated $\omega_0$ 
for nine $v$ values near each of the three tune-out wavelength at 790 nm,
423 nm, and 421 nm. To
each point we added a random experimental error. To account for
variations in experimental sensitivity, we used an error estimate
\be
\delta\omega^2 = 
\frac{\delta\alpha^2 + \delta v^2 
\big[\alpha^{(1)}\big]^2}{(d\alpha/d\omega)^2}
\ee
with $\alpha^{(1)}$ and $d\alpha/d\omega$ calculated numerically
at each measurement point. Here $\delta\alpha$ accounts for the overall
experimental
sensitivity and $\delta v$ accounts for experimental polarization
control. From the $5P$ results in \cite{Leonard15} we estimate
$\delta\alpha = 0.02$ au and $\delta v = 3\times 10^{-6}$. The
resulting $\delta\omega$ values are used as the standard deviation
for a random Gaussian error. This generated a total of 27 simulated
data points $\omega\sub{data}$.

To this data set we fit the solutions of $\alpha^{(0)} - v\alpha^{(1)} = 0$,
using seven adjustable parameters. Three are the matrix element ratios
$R_{5,\,3/2}$, $R_{6,\,1/2}$ and $R_{6,\,3/2}$ where
$R_{nJ} \equiv |D_{nJ}/D_1|^2$. 
Two more parameters are the core polarizabilities
$\left(\alpha_c + \alpha_{cv}^{(0)}\right)/|D_1|^2$ and 
$\alpha_{cv}^{(1)}/|D_1|^2$. The last two parameters
describe the tail contributions.
We define effective matrix elements
\be
|t_{J'}|^2 = \sum_{n'>12} |D_{n'J'}|^2
\ee
so that the scalar tail contribution is
\be
T_{J'} \approx \frac{|t_{J'}|^2}{3} 
\frac{\omega\sub{ion}}{\omega\sub{ion}^2-\omega^2}.
\ee
We normalize these and use fit parameters $|t_{J'}|^2/|D_1|^2$.
For excited states $n' = 7$ to 12, we use the theoretical matrix
elements in Table 1.

\begin{table}
\begin{tabular}{lr}
State & Matrix Element \\ \hline
$6P_{1/2} \rule{0pt}{3ex}  $   &      0.3235(9)   \\
$7P_{1/2}   $   &      0.115(3)    \\
$8P_{1/2}   $   &      0.060(2)    \\
$9P_{1/2}   $   &      0.037(3)    \\
$10P_{1/2}  $   &      0.026(2)    \\
$11P_{1/2}  $   &      0.020(1)    \\
$12P_{1/2}  $   &      0.016(1)    
\end{tabular} \hspace{2em}
\begin{tabular}{lr}
State & Matrix Element \\ \hline
$6P_{3/2} \rule{0pt}{3ex}  $   &      0.5230(8)   \\
$7P_{3/2}   $   &      0.202(4)    \\
$8P_{3/2}   $   &      0.111(3)    \\
$9P_{3/2}   $   &      0.073(5)    \\
$10P_{3/2}  $   &      0.053(4)    \\
$11P_{3/2}  $   &      0.040(3)    \\
$12P_{3/2}  $   &      0.033(2)    
\end{tabular}
\caption{Reduced matrix elements 
$\langle 5S_{1/2} || d || n'P_{J'} \rangle$
for intermediate valence states
used in the tune-out frequency calculations. 
Values for $n' = 6$ are taken from \protect\cite{Herold12}
and other values are theoretical estimates described in
\protect\cite{Leonard15}. Values are reported in atomic units,
with estimated errors listed in parentheses.
}
\end{table}

Values for the fit parameters were obtained by minimizing
$\chi^2 = \sum (\omega\sub{data}-\omega\sub{fit})^2$.
We generated 100 different synthetic data sets using the same model
parameters but different errors. We fit each, and the average value
of the fitted parameters agreed well with the model parameters used.
We report the standard deviation of the fitted parameters as an
estimated error for the procedure. The results are shown in Table 2.
The core and tail parameters have been multiplied by $|D_1|^2$ to
report the physically interesting values.  It can be observed
that the fitting errors are generally lower than current
accuracies, and in particular that definite values for the
core and tail contributions can be obtained. 

To gain a clearer understanding of why measuring multiple tune-out 
wavelengths is effective, we can return to the decomposition of
the polarizability used in Section 2. There, measurements
provided a value for the $B$ parameter and for
a combination of $R$ and $A$. The more complicated dependence on $v$
observed at the 421~nm and 423~nm zeros breaks the degeneracy between 
$A$ and $R$, so that at each of these zero a value for $A$, $B$, and $R$ can
be determined. 
The unknown contributions to $A$ and $B$ are the core
and tail terms, which have distinctly different frequency dependence. 
By relating the measurements near 790~nm and 420~nm, these components can
be distinguished and the core parameters isolated. Since $A$ and $B$ depend
differently on the $J=1/2$ and $J=3/2$ tail parameters, these components
can be distinguished as well.

\begin{table}
\begin{tabular}{lrrr}
Parameter & Model Estimate & Model Error & Fit Error \\ \hline
$R_{5,\,3/2}$ \rule{0pt}{3ex}& 1.99221 & $3\times 10^{-5}$ & $3\times 10^{-5}$ \\
$R_{6,\,1/2}$ & 0.00584 & $3\times 10^{-5}$ & $2\times 10^{-6}$ \\
$R_{6,\,3/2}$ & 0.01526 & $5\times 10^{-5}$ & $5\times 10^{-6}$ \\
$\alpha_c +\alpha_{cv}^{(0)}$ & 8.71 & $9\times 10^{-2}$ & $3\times 10^{-2}$ \\
$\alpha_{cv}^{(1)}$ & -0.04 & $4\times 10^{-2}$ & $9\times 10^{-3}$ \\
$|t_{1/2}|^2$ & 0.009 & $9\times 10^{-3}$ & $1\times 10^{-3}$ \\
$|t_{3/2}|^2$ & 0.03 & $3\times 10^{-2}$ & $3\times 10^{-3}$ 
\end{tabular}
\caption{
Model parameters and fitting results for
measurements
of scalar and vector polarizabilities in Rb atoms near 790 nm,
423 nm, and 421 nm. The parameters are model quantities to be determined
by a fit to experimental data, as explained in Section 3.
The model estimate and model error are the
current best estimates for these parameters. The matrix element
ratios are defined by $R_{n',J'} = |D_{n',J'}/D_1|^2$.
In the case of
$R_{5,3/2}$ the estimate is taken from \protect\cite{Leonard15},
while~$R_{6,J'}$ values are from \protect\cite{Herold12}.
Other estimates are theoretical values described in
\protect\cite{Leonard15} and the current text. The fit error
values are the errors obtained by fitting
synthetic data sets having accuracy comparable to the experimental results
reported in
\protect\cite{Leonard15}. Data values are either dimensionless
or reported in atomic units.
}
\end{table}

\section{Experimental Implementation}

We briefly discuss how an experimental measurement of the type
described might be implemented. A variety of experimental techniques
have been demonstrated for tune-out spectroscopy, including atom interferometry
\cite{Holmgren12,Leonard15}, Bragg coupling in an optical lattice
\cite{Lamporesi10,Herold12,Schmidt16}, or
direct dipole force measurements \cite{Henson15}.
The highest precision has been achieved with the method of
\cite{Leonard15}. Here an atom 
interferometer is implemented in a time-orbiting magnetic
trap potential. 
An off-resonant standing-wave laser is used
to split and recombine the atoms in a Bose-Einstein condensate. 
While the atomic wave packets
are separated, one packet is exposed to a focused traveling-wave
Stark laser beam. This shifts the energy of the atoms according to
\eqref{energy}, which can then be detected as a phase shift in the 
interferometer output. The tune-out wavelength is located by adjusting
the Stark laser frequency such that no phase shift is observed.

This method is intrinsically very sensitive. For the measurements 
proposed here, however, it is also necessary to precisely control
the polarization of the Stark beam. This is difficult to achieve
with purely optical techniques because the vacuum window through
which the beam passes can change the polarization in an uncontrolled way
due to stress-induced birefringence.  In the work of \cite{Leonard15}, we
were able to achieve $v=0$ by alternating the direction of $\bhat$
and thus the sign of $v$.
By ensuring that the interferometer phase did not vary in response,
we could
adjust the polarization to provide $v=0$ at the atoms. Also, during
the actual measurement, the magnetic field establishing $\bhat$ rotated
rapidly so that any residual polarization error was time-averaged
to zero. These methods cannot be directly applied to the measurements
needed here since non-zero values of $v$ are required.

Instead, we propose to apply pure circularly polarized light
to the atoms. This can be established by, for instance, tuning the
Stark laser to the $5P_{1/2}$ resonance. Since atoms in our ground state
cannot scatter $\sigma_+$ polarized light on this transition, the
light polarization can be optimized by minimizing the scattering rate.
The residual scattering rate obtained can serve to characterize the
purity of the polarization.

To this end it is convenient that circularly polarized light is not
highly sensitive to birefringence errors. If the vacuum window or
other optical elements have a small birefringence $\beta$, then the effect on $v$
varies only as $\beta^2$. This can be compared to the linearly
polarized case, where $v$ generally varies linearly with $\beta$ around $v=0$.
From our measurements with linearly polarized light we find
$\beta$ to drift by about $10^{-3}$, so in the circularly-polarized
case stabilities of $10^{-6}$ should be achievable.

Circularly polarized $\sigma_+$ light corresponds to $v = -0.5$.
In order to vary $v$, we propose to vary the magnetic field
direction $\bhat$. In our magnetic trap the bias field rotates
at a frequency $\Omega \approx 2\pi\times 12$~kHz.
If we ensure that the plane of rotation contains the propagation vector
of the Stark beam, then $v$ will vary in time as $-0.5\cos\Omega t$.
The total interferometer phase developed varies as the time average
$\langle v \rangle$.
By applying pulses of the Stark beam that are appropriately synchronized 
with the field rotation, arbitrary values of $\langle v\rangle$
can be achieved. Since timing measurements can be very precise,
accurate control of $\langle v \rangle$ is possible. Initial
alignment in space and time 
between the pulses and the field can be obtained using
the $5P_{1/2}$ resonance as described above, since pure
$\sigma_+$ light can only be obtained when $\bhat$
is exactly parallel to the Stark beam.

We estimate that this technique can provide adequate polarization
control for the measurements suggested here. An alternative
approach would be to use a vacuum window design that
minimizes uncontrolled birefringence \cite{Solmeyer11} and
use conventional optical control to provide the required light polarization.

\section{Conclusions}

On the basis of the analysis discussed here, we conclude that
practical tune-out spectroscopy measurements can provide detailed
information about various components contributing to the
atomic polarizability. We expect that such information  will be 
useful for interpreting atomic parity violation
experiments and similar studies. 

It can be noted the all the parameters discussed here are normalized
to a reference matrix element, in our case $D_1$.
It is easy to see that such normalization is unavoidable, since tune-out
measurements never provide an absolute magnitude for the polarizability,
but only a relationship between different components. In the case of Rb this
imposes a relative error of about $10^{-3}$, which would become
the dominant error for the $n' = 5$ and $n' = 6$ matrix elements
if a measurement result such as in Table 2 were achieved. However,
this would have negligible impact on the core and tail parameters.
The accurate matrix element ratio measurements
achievable with tune-out spectroscopy also mean that any improved
direct measurement of a matrix element could be applied to several other
states. Similar benefits would be obtained from an accurate
measurement of the dc polarizability $\alpha(0)$ \cite{Gregoire15}.

It can also be noted that the measurement technique proposed here
uses a Bose condensate in a magnetic trap. This limits its applicability
in cases where Bose condensation is impossible, or where
interactions make magnetically trapped condensates unfavorable.
However, other measurement techniques might be feasible for such cases.
In particular, Cs measurements would be of interest for more direct application
to existing parity violation results. While Cs is not favorable for 
magnetic trapping \cite{Weber03}, 
it might be possible to achieve similar precision
using an optically trapped condensate interferometer, such as 
demonstrated in \cite{Shin04}.

While extending this type of study to Cs atoms may be a challenge,
we expect that results obtained in atoms like Rb and K would still be
useful in providing guidance for improving theoretical techniques.
Several methods have been developed to calculate, for instance,
tail contributions \cite{Dzuba12b}. Experimental results would allow
such methods to be evaluated, and by using the best method the theoretical
errors for unmeasured atoms like Cs can be improved. Precise
measurements of matrix elements ratios are also useful, since they can
provide a useful benchmark for checking theory calculations \cite{Leonard15}.
It might also be possible to develop semi-empirical methods that
incorporate experimental ratio values, similar to the scaling techniques
that incorporate experimental spectroscopic results for state
energies \cite{Mitroy10}.

\acknowledgments{\textbf{Acknowledgments:} This work was supported by NASA 
(Contract number 1502012). A. Fallon was supported by 
the Jefferson Scholars Foundation. We are grateful to 
Marianna Safronova for discussions and theoretical results.}

\authorcontributions{\textbf{Author Contributions:} 
C. Sackett conceived the idea of the project and wrote the bulk of the 
paper. A. Fallon helped develop the idea, checked the calculations,
helped revise the paper, 
and provided analysis of the expected experimental precision.}

\conflictofinterests{\textbf{Conflicts of Interest:} 
The authors declare no conflict of interest.}

\bibliographystyle{mdpi}
\renewcommand\bibname{References}
\bibliography{sackett}

\end{document}